\documentclass[a4paper,fleqn,usenatbib,useAMS]{mnras}
\usepackage{textcomp}
\usepackage{natbib}
\usepackage{lscape}
\usepackage[section] {placeins}
\usepackage{amsfonts, amssymb, amsmath}
\usepackage{aas_macros}
\usepackage{booktabs, times, varioref ,color, multirow, graphicx, gensymb, caption, diagbox}
\usepackage{tabularx}
\usepackage[T1]{fontenc}
\usepackage{courier}
\usepackage{ae,aecompl} 
\usepackage{ulem}

\include{aas_journals}

\newcommand{\subf}{\textsc{Subfind}}
\newcommand{\aqua}{\textsc{Aquarius}}
\newcommand{\galf}{\textsc{GalForm}}
  
\newcommand{\msun}{{\rm M}_{\sun}}
\newcommand{\lkpc}{{\rm kpc}}

\newcommand\plotone[1]
 {\centering \leavevmode \includegraphics[width={0.99\columnwidth}]{#1}}

\newcommand{\plotside}[1]
 {\centering \leavevmode \includegraphics[width={0.95\textwidth}]{#1}}

\title[Satellites in the inner region of MW]
{The abundance of satellite galaxies in the inner region of $\Lambda$CDM Milky Way sized haloes}
\author[Li et al]
{Ming Li$^{1}$\thanks{E-mail:~mingli@nao.cas.cn}, \,Liang Gao$^{1,2}$, \,Jie Wang$^{1}$ \\
$^1$ Key Laboratory for Computational Astrophysics,\, National Astronomical Observatories, \\
         Chinese Academy of Sciences, Beijing, 100012, China. \\
$^2$ Institute of Computational Cosmology, Department of Physics, University of Durham, Science Laboratories, \\
         South Road, Durham DH1 3LE.}

\date{\today}
\pubyear{2018}

\begin{document}
\label{firstpage}
\pagerange{\pageref{firstpage}--\pageref{lastpage}}
\maketitle

\begin{abstract}
    The concordance $\Lambda$CDM cosmology predicts tens of satellite galaxies
    distributed in the inner region ($<40\ \lkpc$) of the Milky Way (MW), yet
    at most $12$ were discovered at present day, including 3 discovered very
    recently by Dark Energy Survey (DES) and 5 from other surveys (e.g.
    PanSTARRS, MagLiteS). We use $5$ ultra-high resolution simulations of MW
    sized dark matter haloes from the $\aqua$ project, combined with $\galf$
    semi-analytical galaxy formation model, to investigate properties of the
    model satellite galaxy population inside $40\ \lkpc$ of MW sized haloes. On
    average, in each halo this model predicts about $20$ inner satellite
    galaxies, among them $5$ are comparable to the classic satellites in the
    luminosity, these are in stark contrast to the corresponding numbers in
    observations. We further investigate the survivability of these model inner
    satellites in the presence of a central stellar disk with a set of ideal
    simulations. These are done by re-evolving a quarter (30) of the whole
    $\aqua$ inner satellite galaxies (121) by including a static disk potential
    in addition to the MW halo. Our finding is that the additional disk
    completely disrupt 40 percents of these satellites and results in $14$
    satellite galaxies within the $40\ \lkpc$ of each $\aqua$ at the end, in
    reasonably well agreement with observations. 
\end{abstract}



\section{Introduction}
\label{sec:intro}
The $\Lambda$ Cold Dark Matter model (dubbed $\Lambda$CDM) has been extremely
successful to predict various observational properties and the evolution of the
large-scale structure of the Universe. However, it is not equally well to
predict the galaxy properties on the galactic scale and below. For instance,
there has been a long debate on whether the $\Lambda$CDM theory can accommodate
the observed abundance and internal structure of satellite galaxies in our
Milky Way, namely the so-called ``missing
satellite''\citep{1999ApJ...522...82K,1999ApJ...524L..19M}, ``core and
cusp''\citep{2005ApJ...621..757S,2010AdAst2010E...5D,2010MNRAS.408.2364S,2011ApJ...742...20W,2015MNRAS.451.2524M}
and ``too big to fail''\citep{2011MNRAS.415L..40B,2012MNRAS.422.1203B}
problems.

\citet{2010MNRAS.403.1283G} (hereafter G10) put forward a related problem on
this regard. In G10, the authors use a set of ultra-high resolution dark matter
only simulations of Milky Way (MW) sized haloes, and find there are quite
abundant dark matter subhaloes residing in the inner $40\ \lkpc$ of their host
haloes, among these about $20-30$ should be relics of the first galaxies
shining light at present day because they were massive enough to cool by atomic
hydrogen cooling before reionization. On the contrary, among the observed MW
satellite galaxies from SDSS, DES and other surveys (PanSTARRS, MagLiteS)
combined, 12 are within the same distance at present time, including not
conclusively confirmed ones. Hence, the results may point out a discrepancy in
the abundance of satellites in the inner region of the MW between observation
and theory.

The galaxy formation model used in G10 is robust yet simple by using atomic
cooling argument to judge whether or not a halo can form stars, but make no
prediction on properties of the satellite galaxies. In this short paper, we
compensate G10 by taking advantage of the power of a sophisticated galaxy
formation model $\galf$ \citep{2006MNRAS.370..645B,2011MNRAS.417.1260F}, to
make more detailed predictions of properties of inner satellite galaxies in the
MW sized dark matter haloes, and compare with observations to investigate
whether or not the abundance of inner satellite galaxy is a problem of
$\Lambda$CDM Cosmology. Moreover, we will take into account the impact of a
stellar disk in the centre of MW halo on the survivability of these inner
satellite galaxies.  Hydrodynamic simulations from previous studies on this
subject often have a much poorer resolution, we compensate these studies by
performing a sequence of ideal simulations with varying resolutions in order to
carry out numerical convergence study. 

The organization of this paper is as follows. In Section \ref{sec:simu_intro},
we briefly introduce the numerical simulations and galaxy formation model used
in this study. In Section \ref{sec:gal_prop}, we use $\galf$ to predict the
satellite population within $40\ \lkpc$ of MW and compare them with
observations. In Section \ref{sec:ideal_simu}, we present ideal simulations in
order to assess the impact of a stellar disk on the survivability of the model
inner satellite galaxies. In Section \ref{sec:conc}, we summarize our results
and draw conclusions.

\section{The Cosmological simulations}
\label{sec:simu_intro}
Numerical simulations used in this work comprise high resolution re-simulations
of 5 individual MW sized dark matter haloes and their surroundings from the
$\aqua$ Project \citep{2008MNRAS.391.1685S}. These dark matter haloes have
masses in the range $1\sim2\times 10^{12}\ \msun$, comparable to typical values
of our MW. These haloes are randomly selected samples from a large cosmological
simulation, imposing a weak isolation criterion by requiring the candidate halo
to have no companion with the mass greater than half of its own at a distance
less than $1\ h^{-1}\mathrm{Mpc}$ at $z=0$ \citep{2010MNRAS.402...21N}. No
further additional constraints (e.g. the chance to find MW-like haloes with LMC
and SMC liked satellites in a cosmological simulation
\citep{2011ApJ...743...40B,2011ApJ...743..117B}) are applied.

Each $\aqua$ halo has been re-simulated with ``zoom in'' technique with various
resolutions to carry out numerical convergence studies. Here we use the
simulations with level 2 resolution which contains about $10^8$ particles
inside the virial radius of each halo. Five of the six haloes (Aq-A to Aq-E)
are used for our analysis, except for Aq-F which experienced a recent major
merger event at  $z \sim 0.6$ \citep{2015MNRAS.453..377W}.

The $\aqua$ simulation suits assume cosmological parameters as $\Omega_{\rm
m}=0.25$, $\Omega_{\Lambda}=0.75$, $\sigma_8=0.9$, $n_{\rm s}=1$ and $h=0.73$.
These values deviate from the latest Planck results
\citep{2014A&amp;A...571A...1P,2016A&amp;A...594A...1P}, but this small offset
has a negligible effect on our main results.
 
At each recorded snapshot, dark matter haloes are identified with the
friends-of-friends (FoF) algorithm by linking particles separated by 0.2 times
the mean inter-particle separation \citep{1985ApJ...292..371D}. Based upon the
FoF group catalogue, the $\subf$ \citep{2001MNRAS.328..726S} is applied to
identify local over-dense and self-bound dark matter subhaloes; merger trees
are constructed by linking each subhalo at successive snapshots to its unique
descendant using the algorithm described in \citet{2003MNRAS.338..903H}.  We
follow the baryonic evolution using the semi-analytic model $\galf$ developed by
\citet{2011MNRAS.417.1260F}. The model explicitly follows the evolution of the
dark matter halo within which a galaxy forms, and after the halo is accreted to
a larger object and becomes a satellite galaxy. 

Compared to an earlier version of $\galf$ \citep{2006MNRAS.370..645B}, there
are quite a few improvements in \citet{2011MNRAS.417.1260F}, including the use
of a higher yield, a modified supernova feedback model and an earlier
reionization epoch model. The model (fbk:sat/rei:G+L) matches a large body of
observational data on the MW satellite galaxies, whilst the authors did not
carry out a detailed comparison of the inner satellite galaxy population in
observation and their model as we study it here.

\begin{figure}
    \plotone{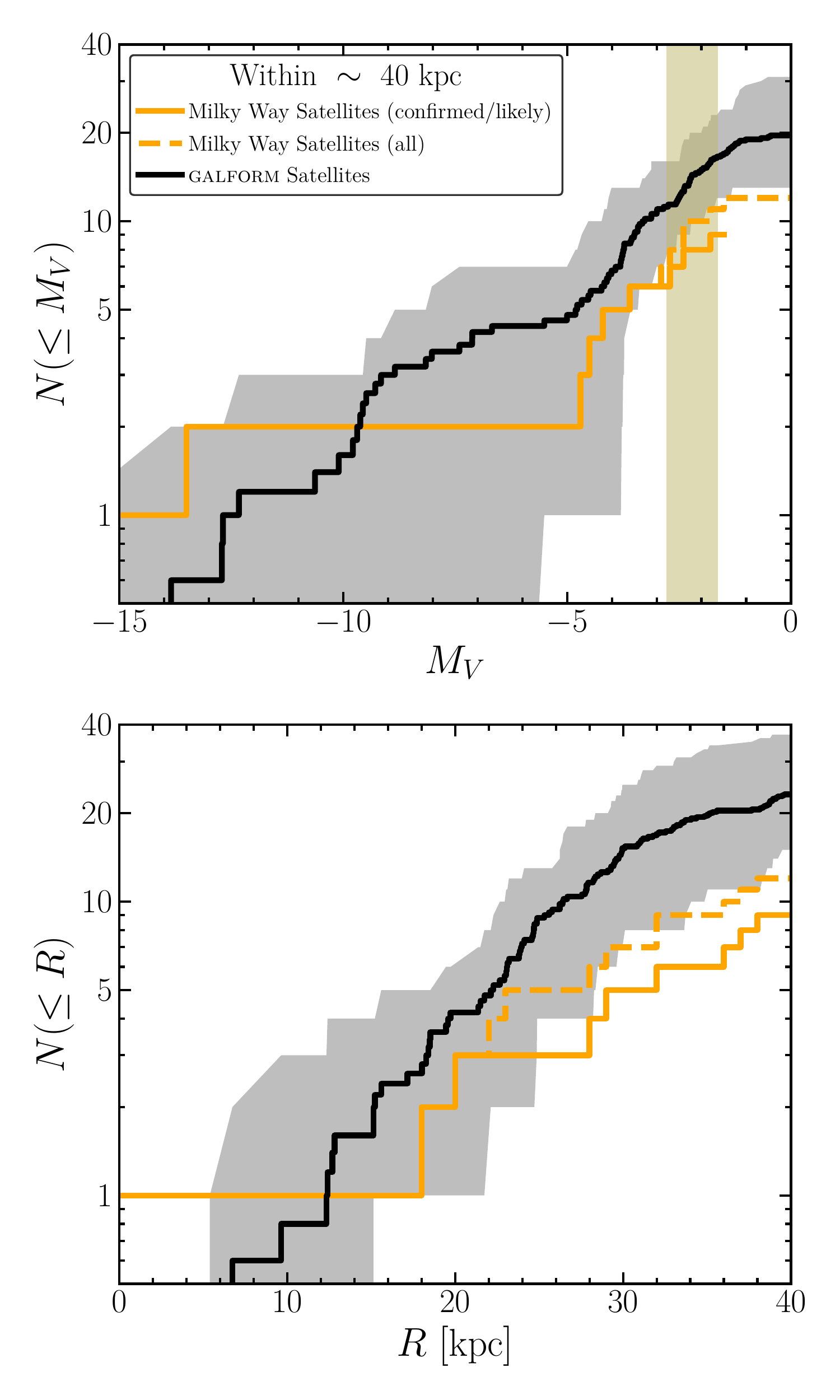}
    \caption{$V-$band luminosity function (top panel) and spacial distribution
    (bottom panel) of the model satellite galaxies within $40\ \lkpc$ of the
    $\aqua$ haloes. The thick black line shows the averaged value of five
    $\aqua$ haloes, while the gray region displays the halo-to-halo variation. The
    thick orange lines show the results for the observed MW satellites which are
    listed in Table \ref{tab:gal_tab}. The solid and dashed orange lines distinguish
    the confirmed (or most likely) and the full satellite samples. The vertical
    spanned region in the top panel indicates the $V-$band detection limit at $40\ \lkpc$ from
    the MW centre, computed with eq. 2 in \citet{2008ApJ...688..277T}.}
\label{fig:cuf_galprop}
\end{figure}
\noindent

\section{The inner satellite galaxies in simulations and observations}
\label{sec:gal_prop}

\begin{table*}
\begin{minipage}{0.6\textwidth}
  \caption{Known MW satellite galaxies distributed within $D_{\mathrm{GC}} \leq
  40\ \lkpc$ of the galactic centre. The first section of satellites are classic or 
  discovered from SDSS survey. The second section of satellites are detected
  by DES. The third section of satellites are detected from other surveys. In
  the table, the distance of each satellite galaxy from the MW centre (the sun)
  $D_{\mathrm{GC}}$ ($D_{\sun}$), $V-$band magnitude $M_V$, stellar mass
  $M_{\star}$ and dynamical mass within half light radius $M_{\rm
  dyn}(\leqslant r_{\rm h})$ (if available) are listed.}
  \label{tab:gal_tab}
  \begin{center}
  \begin{tabular}{@{}l|cccc}
  \toprule
      Name & $D_{\mathrm{GC}} / D_{\sun}$ & $M_{V}$ & $M_{\bigstar}$ & $M_{\rm dyn}(\leqslant r_{\rm h})$\\ 
           & $(\lkpc)$                    & (mag) & $(\msun)$        & $(\msun)$\\
  \midrule
   & \multicolumn{4}{c}{Classic / SDSS$^a$} \\
  Sagittarius dSph       & 18.0 / 26.0 & -13.5 &  $2.1\times10^7$ & $1.9\times10^8$\\ 
  Segue I                     & 28.0 / 23.0 &  -1.5 &  $3.4\times10^2$ & $2.6\times10^5$\\ 
  Ursa Major II             & 38.0 / 32.0 &  -4.2 &  $4.1\times10^3$ & $3.9\times10^6$\\ 
  Bootes II                   & 40.0 / 42.0 &  -2.7 &  $1.0\times10^3$ & $3.3\times10^6$\\ 
  \midrule
     & \multicolumn{4}{c}{DES} \\
  Tucana III (DES J2356-5935)$^b\ ^{\dag}$   & 23.0 / 25.0 & -2.4  &  $8.0\times10^2$ & --- \\
  Ret II (DES J0335.6-5403)    $^{b,c}$    & 32.0 / 30.0 & -3.6  &  $2.6\times10^3$ & $2.4\times10^5$ \\
  Cetus II (DES J0117-1725)   $^b\ ^{\dag}$    & 32.0 / 30.0 &  0.0  &  $1.0\times10^2$ & --- \\
  \midrule
     & \multicolumn{4}{c}{DECam} \\
  Hydrus I$^d$          & 20.0 / 27.6 &  -4.7  &  $6.0\times10^3$ & $2.6\times10^5$ \\   
     & \multicolumn{4}{c}{Pan-STARRS} \\
  Draco II        $^e\ ^{\dag}$   & 22.0 / 20.0 &  -2.9  &  --- & --- \\ 
  Triangulum II$^f$   & 36.0 / 30.0 &  -1.8  &  --- & --- \\ 
     & \multicolumn{4}{c}{MagLiteS} \\
  Carina III$^g$         & 29.0 / 27.8 &  -2.4  &  --- & --- \\ 
  Carina II $^g$         & 37.0 / 36.2 &  -4.5  &  --- & --- \\ 
  \bottomrule
\end{tabular}
\medskip
\end{center}

  $^a$ \citet{2012AJ....144....4M}, $^b$ \citet{2015ApJ...813..109D}, $^c$ \citet{2015ApJ...811...62K}, $^d$ \citet{Koposov_2018}, 
  $^e$ \citet{Laevens_2015b}, $^f$ \citet{Laevens_2015a}, $^g$ \citet{Torrealba_2018}
  
  $^{\dag}$  Dwarf galaxy whose identity has not been conclusively confirmed (private communication with Josh Simon).
  

\end{minipage}
\end{table*}

In Figure \ref{fig:cuf_galprop} we present the cumulative $V-$ band luminosity
function and the spatial distribution of the model satellite galaxies within
$40\ \lkpc$ of five $\aqua$ haloes. The dark solid line shows the averaged
count and the light shaded area displays the whole scatter of five haloes. We
also show results for the known satellite galaxies within the same distance in
the same figure with orange solid line. These known satellite galaxies are
collected from SDSS \citep[Table1-3]{2012AJ....144....4M}, DES
\citep{2015ApJ...813..109D,2015ApJ...811...62K,Koposov_2018}, PanSTARRS 
\citep{Laevens_2015a,Laevens_2015b}, and MagLiteS \citep{Torrealba_2018}.  We list the
properties of these satellite galaxies in Table \ref{tab:gal_tab}. Note, 3 of
them are not conclusively confirmed as dwarf galaxies. Including these 3, the
luminosity function of the observed inner satellite galaxies is shown as orange
dashed line. The vertical yellow shaded area indicates the observational
detect limits of satellite galaxy of the SDSS survey
\citep{2008ApJ...686..279K,2009ApJ...696.2179K,2008ApJ...688..277T}.

Apparently, most of these inner model satellite galaxies are detectable with
SDSS survey. On average, our simulations predict about 20 satellite galaxies
within $40\ \lkpc$ of the MW, in agreement with the simple model of G10.  $5$ of
them are as bright as classic satellites identified in observations with
$M_V<-5$. Comparing with observations, our simulations predict 2 times more
inner satellite galaxies than the known ones in current observations. In
particular, our model predicts about 2 times more satellite galaxies more bright
than $M_V<-5$.

The cumulative spatial distribution of these model and observational inner
satellite galaxies are presented in the lower panel of the same figure.
Comparing the two, they only roughly agree with each other within $20\ \lkpc$. 
But the model predicts much more satellite galaxies beyond it. Note, almost no
satellite galaxies are founded within this distance in the latest highest
resolution hydrodynamic simulations of the MW, e.g. the $\textsc{APOSTLE}$
project \citep{2016MNRAS.457.1931S,2017MNRAS.467.4383S} and the
$\textsc{Latte}$ simulation \citep{2016ApJ...827L..23W,2017MNRAS.471.1709G}. 
In the later section, we will show that this is very likely due to a poor
numerical resolution of these simulations.

\begin{figure*}
\plotside{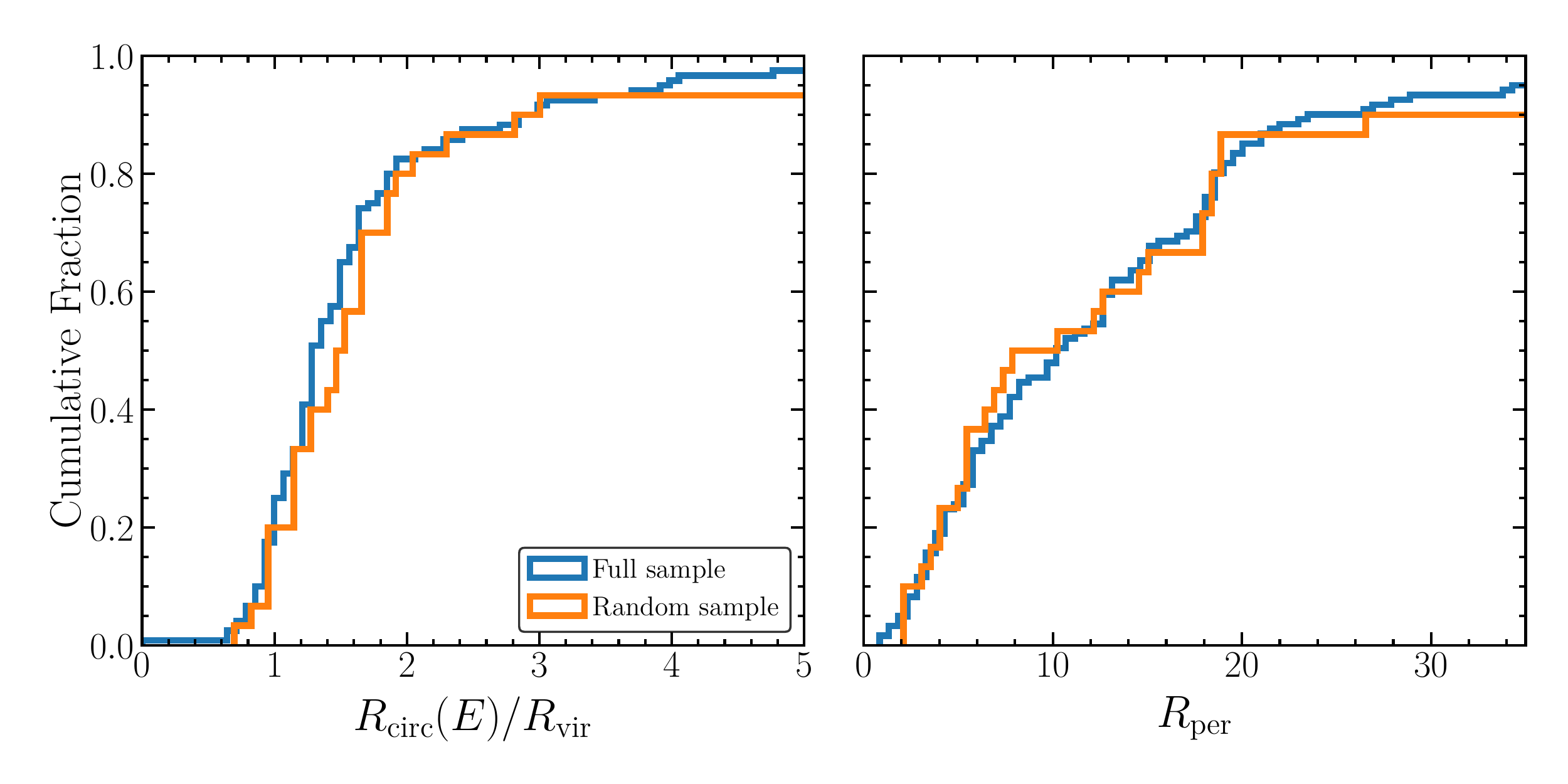}
\caption{Cumulative distribution functions of orbital energy parameter $R_{\rm
    circ}(E)/R_{\rm vir}$ at infall (left) and the first pericentric radius
    $R_{\rm per}$ since infall (right) of the $\aqua$ inner satellite galaxies. The
    blue solid lines are for the full sample, and the orange solid lines are
    for the randomly selected 30 satellite galaxies.}
\label{fig:infall_orbit_prop}
\end{figure*}

The $\aqua$ haloes are simulated with dark matter only, without considering 
the central gaiant disk like MW. The presence of a disk may have a substantial impact
on the abundance of satellite galaxies, especially in the inner region at which
we are discussing in this paper. The impact of a disk on the abundance of inner
satellite galaxies has been investigated in previous studies 
\citep[e.g.][]{2010ApJ...709.1138D,2015MNRAS.452.2367Y,2017MNRAS.465L..59E,2016arXiv161207834J,2017MNRAS.471.1709G}.
We compensate these studies in the following sections by performing a series of
much higher resolution numerical experiments. 

Before resorting to the numerical experiments, we first make a rough assessment
of the impact of disk on the tidal disruption of these model inner satellite
galaxies by examining their orbit parameter distribution. If their pericentric
radius is within the size of the MW disk, the impact due to disk is expected
to be strong and vice versa \citep{2017arXiv171204467N,2017MNRAS.465L..59E}.  In
Figure \ref{fig:infall_orbit_prop}, we present the cumulative distribution function of the
orbital properties for the inner satellite galaxies in $\aqua$ simulations.
Here we characterise the orbit properties with the first pericentric radius
$R_{\rm per}$ since infall, and $R_{\rm circ}(E)/R_{\rm vir}$, the radius of
the circular orbit corresponding to the orbital energy of the satellite at the time
of infall, expressed in virial radius of the host halo.  \footnote{Defined as
the radius within which the mean density is 200 times of the critical density
at the time of infall.}

Clearly, about 50 per cent of satellite galaxies have their first pericentric
radius $R_{\rm per}$ within $10\ \lkpc$, indicating that these satellites are
prone to be affected by the disk. Apart from the pericentre, orbital energy is
also an important parameter to the tidal disruption of satellite galaxy, we
show the orbital energy  distribution of our whole sample in the right-hand
panel of the same figure.

\section{The impact of a stellar disk on the abundance of inner satellite galaxies}
\label{sec:ideal_simu}
\subsection{Models and numerical experiments}
\label{sec:simu_setup}

\textit{Model of the MW.}
The MW comprises a dark matter halo, a disk and a bulge. For
simplicity, we neglect the bulge component and treat the dark matter halo and
disk components as rigid background potentials and thus neglect the effect of
dynamical friction. We assume that dark matter distribution of the MW follows the 
NFW \citep{1996ApJ...462..563N} profile. The potential corresponding to a
NFW halo reads
\begin{equation}
  \Phi_{\rm NFW}(R)=-\frac{G\ M_{\rm vir}}{R}\frac{\ln(1+C_{\rm vir}\ R/R_{\rm
    vir})}{\ln(1+C_{\rm vir})-/(1+C_{\rm vir})},
\label{eq:phi_nfw}
\end{equation}
$R_{\rm vir}$ is defined as the virial radius within which the
mean over-density is 200 times of the critical density and the corresponding
mass within it is defined as the virial mass $M_{\rm vir}$. In this work, the
MW sized halo is modelled with $M_{\rm vir} = 10^{12}\ \msun$, $R_{\rm vir} =
160.6\ \lkpc$ and concentration $C_{\rm vir}=12$. These parameters are chosen
according to previous works \citep{2004ApJ...601...37K,2010MNRAS.406.1290P,2012MNRAS.427.1429W}.

We model the disk as an embedded potential in cylindrical coordinates which 
follows an axisymmetric disk model \citep{1975PASJ...27..533M}:
\begin{equation}
  \Phi_{\rm d}(R_{\rm XY}, Z)=-\frac{G M_d }{\sqrt{R_{\rm XY}^2 + (a+\sqrt{Z^2+b^2})^2}},
\label{eq:phi_d}
\end{equation}
with radial and vertical scale lengths $a = 6.5\ \lkpc$
and $b = 0.25\ \lkpc$, and mass $M_{\rm d} = 0.1 M_{\rm vir}$.

\noindent
\textit{Model of satellite galaxy.}
The orbiting satellite galaxy is discretized with N-body model by generating an
equilibrium particle realisation, the density distribution follows a truncated
NFW profile \citep{2004ApJ...608..663K}
\begin{displaymath}
\label{eq:rho_sub}
\rho(r)=\left\{
\begin{array}{ll}
    \frac{\rho_0}{(c_{\rm vir}r/r_{\rm vir})^\gamma (1+(c_{\rm
    vir}r/r_{\rm vir})^\alpha)^{(\beta-\gamma)/\alpha}} & \textrm{$(r\leqslant r_{\rm vir})$}, \\
    
    \frac{\rho_0}{c_{\rm vir}^\gamma (1+c_{\rm
    vir}^\alpha)^{(\beta-\gamma)/\alpha}}\bigg(\frac{r}{r_{\rm
    vir}}\bigg)^\varepsilon \exp\bigg(-\frac{r-r_{\rm vir}}{r_{\rm
    dec}}\bigg) & \textrm{$(r>r_{\rm vir})$}. 
\end{array}
\right.
\end{displaymath}

\noindent
$\rho_0$ and $c_{\rm vir}$ are the characteristic density and the concentration
of the satellite respectively.  $r_{\rm dec}$ is a parameter that controls the
sharpness of the slope transition towards an exponential cutoff and is set to
0.3 times of satellite's virial radius $r_{\rm vir}$. To obtain a continuous
logarithmic slope, $\varepsilon$ is defined as
\begin{eqnarray}
\varepsilon = \frac{-\gamma -\beta c_{\rm vir}^\alpha}{1+c_{\rm vir}^\alpha} +
    \frac{r_{\rm vir}}{r_{\rm dec}}.
\label{eq:eps}
\end{eqnarray}
Throughout this work, we adopt a cuspy NFW density profile with $(\alpha, \beta, \gamma)=(1,3,1)$. 

\begin{table}
    \caption{Overview of the detailed properties of the satellite galaxies used
    in our resolution test. Columns $(2-4)$ show the virial mass, radius and
    concentration (calculated with the halo mass-concentration relation from
    \citet{2008MNRAS.390L..64D}) respectively. Column $(5-6)$ are particle
    numbers and the corresponding softening lengths used for the resolution
    test and ordinary runs. } \label{tab:simu_tab}
  \begin{tabular}{@{}cccccc}
  \toprule
   Run              & $m_{\mathrm{vir}}$  
                        & $r_{\mathrm{vir}}$
                        & \multirow{2}{*}{$c_{\mathrm{vir}}$}
                        & \multirow{2}{*}{$N_{\mathrm{part}}$} 
                        & $\epsilon$ \\

                        & $(\msun)$         
                        & $(\lkpc)$
                        & & & $(\lkpc)$\\
  \multicolumn{6}{c}{\vspace{-2mm}}\\
   (1) & (2) & (3) & (4) & (5) & (6) \\
  \midrule                                                                                                             
  \multirow{5}{*}{halo.m1e8} & \multirow{5}{*}{$10^{8}$}  & \multirow{5}{*}{7.548} & \multirow{5}{*}{13.127} & $10^4$ & 0.34  \\
    &  &  &  & $10^5$ & 0.10  \\
    &  &  &  & $10^6$ & 0.03  \\
    &  &  &  & $10^7$ & 0.01  \\                                                                                                      
  \bottomrule
\end{tabular}
\end{table}

The coordinate system is centred on the MW halo. The disk potential is fixed on
the $X-Y$ plane. The N-body simulations presented in this study were carried
out with the P-GADGET3 code \citep{2005MNRAS.364.1105S} under isolated boundary
conditions.

\begin{table}
    \caption{The collection of orbital parameter setups used in our resolution test. 
    Orbital energy $R_{\rm circ}(E)/R_{\rm vir}$, the
    pericentric radius $R_{\rm per}$ and the initial velocity expressed in circular
    velocity at virial radius of the host halo are presented. }
  \label{tab:orbit_tab}
  \begin{tabular}{@{}cccc}
  \toprule

   Run          & $R_{\rm circ}(E)/R_{\rm vir}$
                    & $R_{\mathrm{per}}$
                    & $(v_{x0,} v_{y0}, v_{z0})\ (V_{\mathrm{vir}})$ \\
  \multicolumn{2}{c}{\vspace{-2mm}}\\
  \midrule
  Orbit I & \multirow{2}{*}{1.34} & 16.26 & (-1.19,0,0.04) \\
  Orbit II & & 3.25 & (-1.18,0,0.14) \\
  \bottomrule
\end{tabular}
\end{table}

\begin{figure*}
\plotside{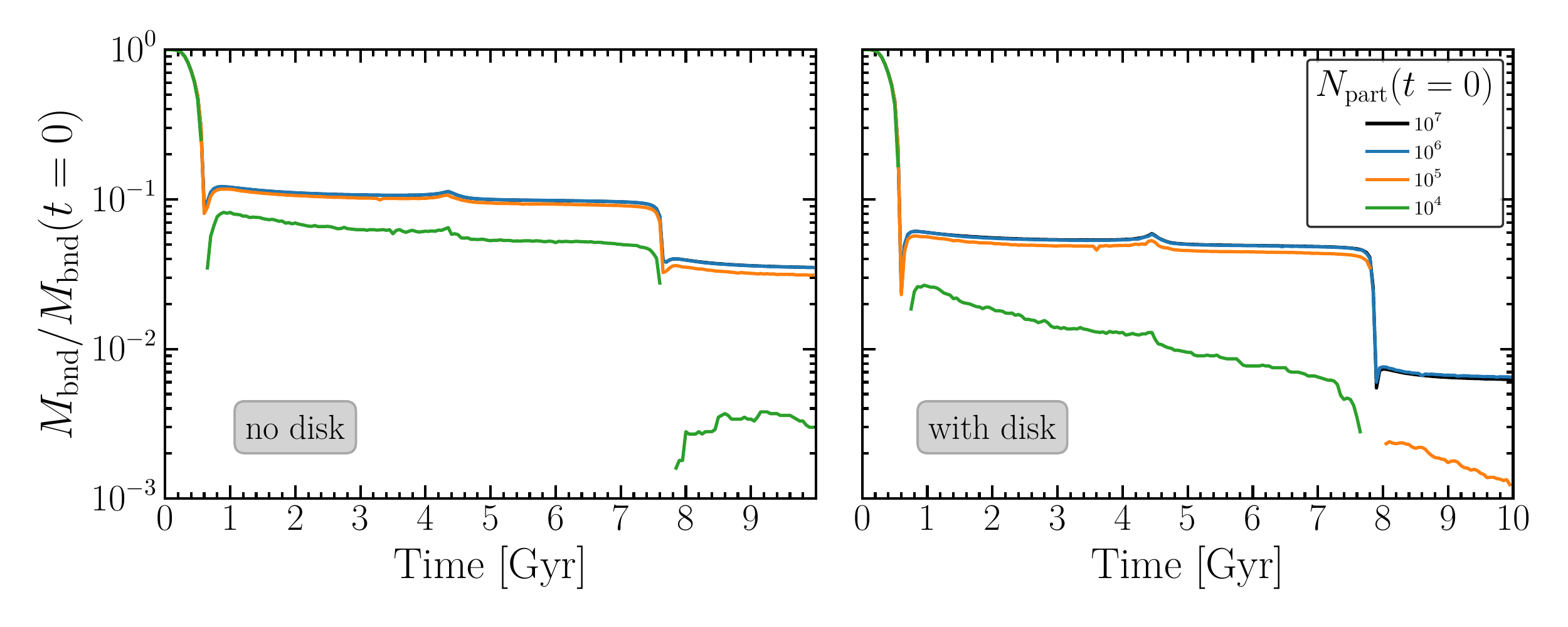}
\plotside{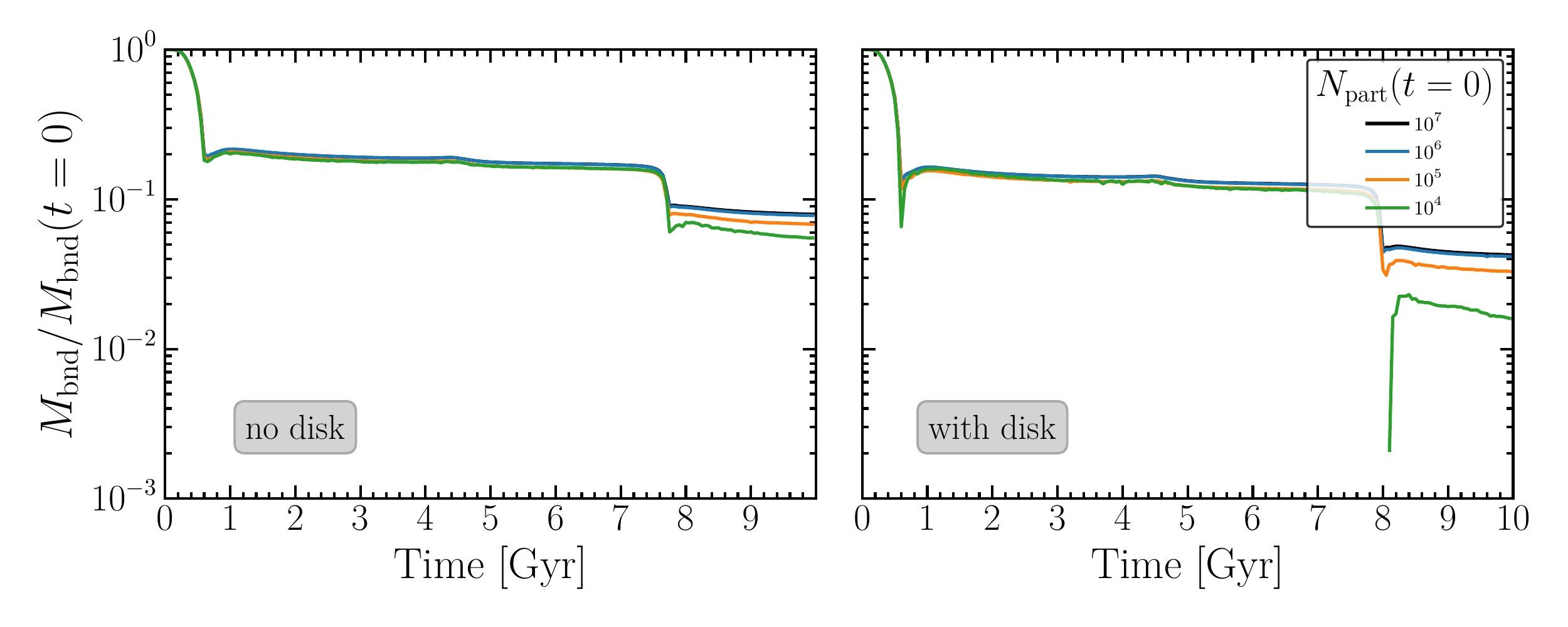}
\caption{The evolution of the bound mass fraction of a satellite galaxy with an
initial mass $10^8\ \msun$ in the MW model without (left) and with (right) a
disc component.  The top and bottom rows show the results for Orbit I and Orbit
II, respectively. Lines with different colours distinguish simulations with
different resolutions.}
\label{fig:mass_reso}
\end{figure*}

\subsection{Resolution Test}
\label{sec:reso_test}
It is important to first identify the required numerical resolution to
reliably resolve the dynamics of satellite galaxies. To this end, we evolve a
satellite galaxy at varying numerical resolutions and assume two sets of
orbital parameters. The mass and orbital energy parameter of the satellite are
assumed to be $m_{\mathrm{vir}}=10^8\ \msun$ and $R_{\rm circ}(E)/R_{\rm vir}=1.34$,
respectively. These parameters are chosen to closely match the typical values
of the satellite galaxies in the $\aqua$ simulation suits. We adopt two
pericentres for the tests by choosing values of $R_{\rm per}$ to be 10 and 50
percentiles of the distribution function shown in the right panel of Figure
\ref{fig:infall_orbit_prop}, representing an extreme and a typical case to
examine the impact of the disk on the tidal disruption of the galaxy. 

We assume the total mass profile of the galaxy following a NFW profile.  Note,
some observational results suggest that MW satellite galaxies may have cored
profiles \citep[e.g.][]{2007ApJ...663..948G,2014ApJ...789...63A,2015AJ....149..180O}. 
As shown by \citet{2010MNRAS.406.1290P} and \citet{2017MNRAS.465L..59E}, the
survivability of satellite galaxy in numerical simulations does depend on the
inner profile of their dark matter halo. Galaxies with cored inner density
profile are more prone to be disrupted than their cuspy counterparts. However,
some recent theoretical works \citep{2016arxiv160706479F,2018MNRAS.474.1398G}
argued that the results of the cored profiles suggested in observation are not
convincing. Hence we adopt the cuspy NFW profile in this study for simplicity.
The concentration parameter of the halo is set to be $c_{\rm vir}=13.1$,
corresponding to the value estimated by the halo mass-concentration relation
given by \citet{2008MNRAS.390L..64D}.

The model galaxies are starting from the virial radius of the MW halo with the
coordinates $(x,y,z)=(1,0,0)\ R_{\rm vir}$, and are evolved for 10 Gyrs which
corresponds to the typical infall redshift $z=2$ of satellite galaxies in the
$\aqua$ simulation suits.

We carried out numerical experiments with different particle numbers, $N_{\rm
part}=10^4, 10^5, 10^6$ and $10^7$. For the galaxy, we follow its evolution in
the MW halo model with and without a disk component under two sets of orbital
parameters. $\subf$ \citep{2001MNRAS.328..726S} is applied to calculate the
residual bound mass of each galaxy.  

In Figure \ref{fig:mass_reso}, we present the evolution of the bound mass
fraction of the satellite galaxy with different resolutions and different
orbital parameters. Upper panels show results for the extreme orbit case Orbit
I and bottom panels are for the typical one, Orbit II. Simulations excluding
and including the disk are shown in the left-hand and right-hand panels,
respectively. In both cases, the numerical resolution has a large effect on the
tidal distribution of the satellite galaxy. Using a number of particles like
$10^4$, as similar to the most up-to-date highest resolution hydrodynamic
simulations, severely underestimate the survivability of the galaxy, 
particularly in the cases including the disk. For the extreme orbital parameter,
our numerical experiments converge at a particle number $10^6$. The particle
number for the convergence is less for the typical case of Orbit II, which is
about $10^5$. From a conservative consideration, in the follows, we will
perform all our experiments with a particle number $10^6$ for each satellite
galaxy. This number is compatible with that used in previous works
\citep{2013MNRAS.431.3533C,2017MNRAS.465L..59E,2017MNRAS.472.3378F} and is much
larger than the highest resolution hydrodynamic simulation in the community.
This should partially account for the fact that these simulations nearly have
no satellite galaxies near the centre. 

\subsection{The impact of the MW disk on the abundance of the inner satellite galaxies}
\label{sec:sub_evol}
\begin{figure}
\plotone{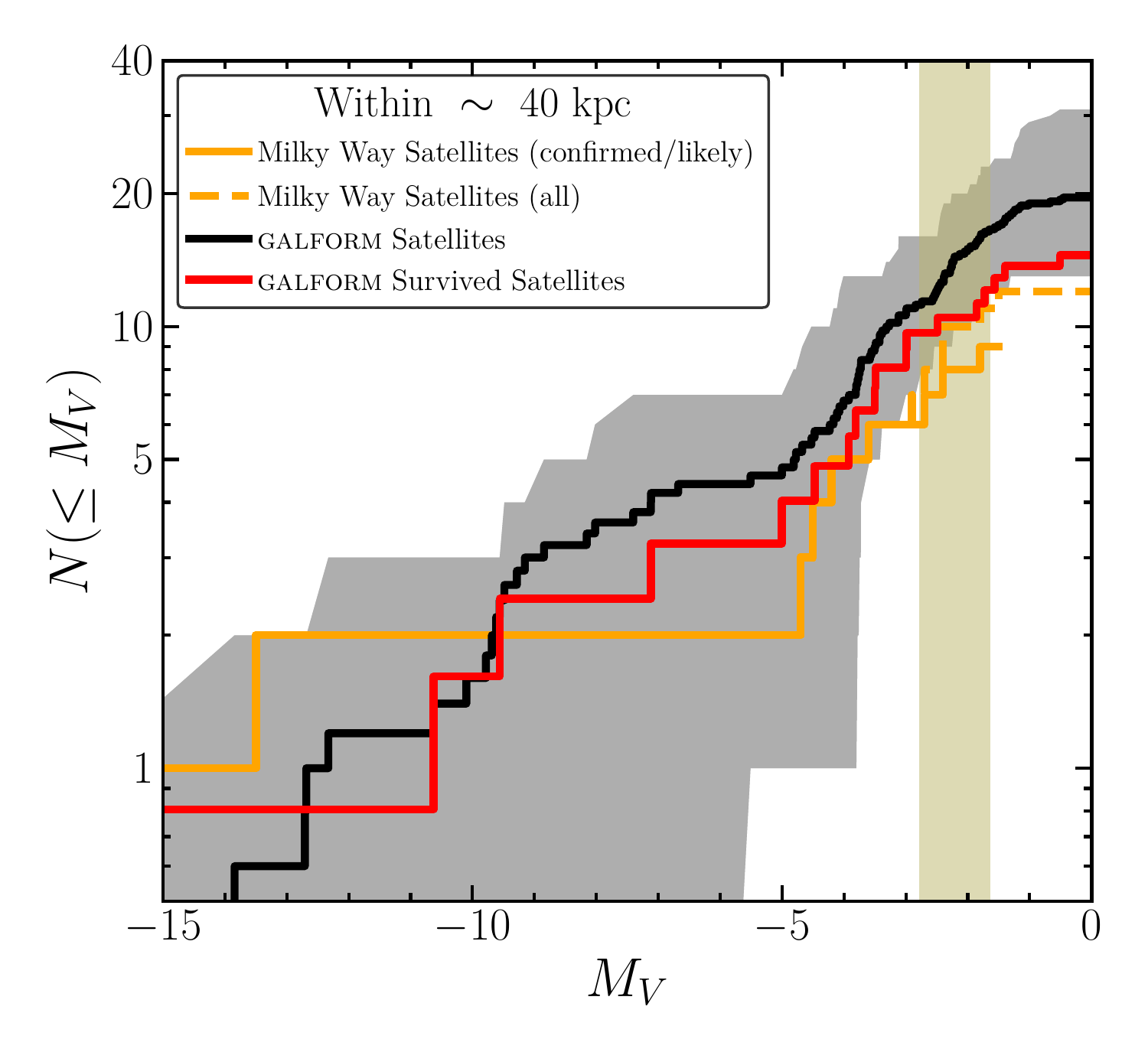}
\caption{The same as the top panel in Figure \ref{fig:cuf_galprop}, but
    adding a curve showing the luminosity function of the inner model 
    satellites after taking into account of the effect of the MW disk.}
\label{fig:sur_galprop}
\end{figure}

To explore the impact of the disk on the tidal disruption of the model inner
satellite galaxies, we randomly select 30 galaxies from our full 121 inner
satellite galaxy sample. The orange lines in Figure \ref{fig:infall_orbit_prop}
display distributions of orbital parameters of this sub-sample. As can be seen,
they agree very well with the whole sample, suggesting that they are a fair
representative sample of the $\aqua$ inner satellite galaxies. For each galaxy
in the randomly selected sample, we follow its evolution from its infall time
with $10^6$ particles in the MW model with the disk we described in the
previous section. The mass distribution of each galaxy is assumed to follow a
NFW profile. Its mass, orbital parameters, concentration parameter and position
are set to be the corresponding values at infall extracted from the $\aqua$
simulation suits. The disk is fixed on the $X-Y$ plane. We also re-run 10 of
these galaxies by varying the disk plane to be $X-Z$ and $Y-Z$, and find the
results hardly change. In the final outcome of our 30 simulations, 12 (40\%)
galaxies are completely disrupted due to the presence of the disk.  Here we
define a galaxy is completely disrupted when $\subf$ is not able to find more
than 32 bound particles. Applying the result to the whole $\aqua$ inner galaxy
sample, 73 out of 121 model inner satellite galaxies should survive to the
present day. Namely on average, each $\aqua$ halo contains 14 inner satellite
galaxies after taking into account the effect of the disk. We present the
corrected cumulative $V-$ band luminosity function of the model satellite
galaxy in Figure \ref{fig:sur_galprop}, it agrees with observations reasonably
well when considering the scatter among 5 $\aqua$ haloes.

\section{Conclusions}
\label{sec:conc}
In this work we make use of the $\aqua$ project--a set of ultra-high resolution
simulations of MW sized dark matter haloes, combined with a sophisticated
semi-analytical galaxy formation model-$\galf$, to investigate the abundance of
satellite galaxies residing within $40\ \lkpc$ of halo centre. Using a simple
atomic cooling argument, G10 suggested that the abundance of MW inner satellite
galaxies may be incompatible with observations. We use $\galf$ to predict
properties of these $\Lambda$CDM model inner satellite galaxies. On average,
about 20 satellite galaxies reside within $40\ \lkpc$ of each $\aqua$ halo,
about a factor of $2$ times exceeding the observed number. Most of these model
inner satellite galaxies are brighter than the detection limit of SDSS survey,
5 of them are as bright as classic satellite galaxies.

Given the apparent inconsistency between the $\Lambda$CDM prediction and
observations, we perform a series of numerical experiments to examine the
impact of the disk on the abundance of the inner model satellite galaxies. To
this end, we randomly select a quarter of the $\aqua$ inner satellite galaxies.
For each of them, we follow its evolution from the infall time to the present
day, with the orbital parameters and positions taken from the original Aquarius
simulation suits. Our finding is that the MW disk has a strong effect to
disrupt satellite galaxies with very close pericentric parameters. As a result,
in the presence of the disk, the number of the model inner satellite galaxies
can be reduced by $40$ per cent when compared with the case without the disk.
For each $\aqua$ halo, the model predicts $14$ satellite galaxies within $40\
\lkpc$, in reasonable well agreement with observations.

Note, when we evolve each galaxy in the simulation, we assume static potentials
for the MW halo and the disk with present-day values during its entire evolution.
This neglects facts that the MW only acquires a fraction of its present-day
mass by then and the MW disk may grow significantly after the infall of the
satellite galaxy. Hence, our results may overestimate the impact of the
disk on the disruption of the inner satellite galaxies. On the other hand, we
may underestimate the tidal disruption of satellite galaxies due to baryonic
effect as discussed in \citet{2017MNRAS.471.1709G}. However, the strong
impact on the disruption of satellites by disk shown in this work certainly
reduces the number of the model inner satellite galaxies by a large factor,
which significantly relieves the large discrepancy between the theory and
observations. A fully convincing work on this requires an ultra-high resolution
and realistic hydrodynamic simulation of the MW galaxy. According to our
numerical experiments, at least $10^5$ particle is required to follow each
satellite galaxies, thus in order to resolve a typical MW inner dwarf galaxy
with a mass $10^8\ \msun$, a dark matter particle mass resolution $1000\ \msun$
is required, far beyond the highest resolution achieved at present day.


\section*{Acknowledgement}
We appreciate the help of the anonymous referee to improve this manuscript.  We
thank Andrew Cooper for providing us $\galf$ galaxy catalogue of the $\aqua$
simulation suits. We are grateful to Josh Simon for clarifying the abundance of
the inner MW satellite galaxies. We are also grateful to Juntai Shen for useful
discussions. We acknowledge support from the National Key Program for Science
and Technology Research and Development (2015CB857005,2017YFB0203300) and NSFC
grants (11390372,11425312,11503032,11573031,11851301 and 11873051). ML also
acknowledges support from CPSF-CAS joint Foundation for Excellent Postdoctoral
Fellows No.2015LH0014. 


\bibliographystyle{mnras}
\bibliography{ref}


\label{lastpage}
\end{document}